\documentclass[10pt,final,conference,letterpaper,twocolumn]{IEEEtran}
\usepackage[bookmarks=false]{hyperref} 
\usepackage{flushend} 
\usepackage[utf8]{inputenc}
\usepackage[english]{babel}
\usepackage{xcolor,color,colortbl,graphicx,tabularx}
\usepackage{amssymb , pifont}
\usepackage{etoolbox}
\usepackage[para]{threeparttable}
\usepackage{pgfplots}
\usepackage{multirow}
\usepackage{xspace}
\usepackage{cite}
\usepackage[acronym]{glossaries}
\usepackage{booktabs}
\usepackage{comment}
\usepackage{subcaption}
\usepackage{subfloat}
\usepackage{floatrow}
\usepackage{todonotes}
\presetkeys%
{todonotes}%
{inline,}{}
\usepackage{nicefrac}

\pdftrailerid{test}

\usepackage{tikz}
\usetikzlibrary{shapes,arrows,shadows}
\usepackage{amsmath,bm,times}
\newcommand\numberthis{\addtocounter{equation}{1}\tag{\theequation}}
\usepackage{environ}
\usepackage{tumcolor}
\usepackage{algorithm,algpseudocode,algorithmicx}
\algnewcommand\algorithmicforeach{\textbf{for each}}
\algdef{S}[FOR]{ForEach}[1]{\algorithmicforeach\ #1\ \algorithmicdo}%
\newlength\myindent
\newcommand{\eg}{\textit{e.g.,}\xspace}
\newcommand{\ie}{\textit{i.e.,}\xspace}

\newcommand{\gt}{ground truth\xspace}
\newcommand{\GT}{Ground Truth\xspace}
\newcommand{\ls}{large-scale\xspace}
\newcommand{\ds}{data set\xspace}
\newcommand{\dss}{data sets\xspace}
\newcommand{\DS}{Data Set\xspace}
\newcommand{\nl}{non-linear\xspace}
\newcommand{\NL}{Non-Linear\xspace}

\colorlet{tablelight}{black!15}
\colorlet{tabledark}{black!40}

\title{Large-Scale Classification of IPv6-IPv4 Siblings \\with Variable Clock Skew}

\author{\IEEEauthorblockN{Quirin Scheitle, Oliver Gasser, Minoo Rouhi, Georg Carle}
		Chair of Network Architectures and Services\\
	\IEEEauthorblockA{Technical University of Munich (TUM)\\
		Email: \{scheitle,gasser,rouhi,carle\}@net.in.tum.de}
}

\hypersetup{
	pdftitle={Large-Scale Classification of IPv6-IPv4 Siblings \\with Variable Clock Skew},
	pdfauthor={Quirin Scheitle, Oliver Gasser, Minoo Rouhi, Georg Carle},
}

\begin{document}

\maketitle
\begin{abstract}%
Linking the growing IPv6 deployment to existing IPv4 addresses is an interesting field of research, be it for network forensics, structural  analysis, or reconnaissance. 
In this work, we focus on classifying pairs of server IPv6 and IPv4 addresses as \textit{siblings}, \ie running on the same machine. 
Our methodology leverages active measurements of TCP timestamps and other network  characteristics, which we measure against a diverse ground truth of 682 hosts.
We define and extract a set of features, including estimation of variable (opposed to constant) remote clock skew. 
On these features, we train a manually crafted algorithm as well as a machine-learned decision tree. 
By conducting several measurement runs and training in cross-validation rounds, we aim to create models that generalize well and do not overfit our training data.
We find both models to exceed 99\% precision in train and test performance. 
We validate scalability by classifying 149k siblings in a large-scale measurement of 371k sibling candidates. 
We argue that this methodology, thoroughly cross-validated and likely to generalize well, can aid  comparative studies of IPv6 and IPv4 behavior in the Internet.
Striving for applicability and replicability,  we release ready-to-use source code and raw data from our study.
\end{abstract}
\section{Introduction}
The emergence of IPv6 in the Internet offers interesting possibilities for studies to compare IPv6 and IPv4 structures and attributes in the Internet.
Interesting questions are, \eg whether and to what extent IPv6 addresses are co-deployed on existing IPv4 hardware, whether correlated IPv6 and IPv4 attacks originate from the same hosts, what levels of redundancy can be achieved by co-deploying IPv6, or to conduct IPv6-IPv4 geolocation comparisons.
An important prerequisite for such studies is the identification and classification of related IPv6 and IPv4 addresses.
One such association can be gained from DNS queries, which yields IPv6-IPv4 address pairs that deliver the same service, but may be hosted on different machines.  
We address the problem of determining whether a set of IPv6 and IPv4 addresses are located on the same machine in a dual-stack setup. 
As in prior work\,\cite{beverly2015server}, we use the term \textit{sibling} for such a relation. 
This level of relation may help to draw deeper conclusions from service-level IPv6-IPv4 comparative studies, \eg on latency\,\cite{bajpai2015ipv4} or security comparisons\,\cite{czyzbackdoor}.
We base our classification approach on active measurements of TCP timestamps, based on prior work by Kohno\,\cite{kohno2005remote}, Zander\,\cite{zander2008improved}, and Beverly and Berger\,\cite{beverly2015server}.
Our approach leverages novel features, such as the identification of unique nonlinear patterns caused by variable skew.
Based on these features, we train and test various classifier models, using thorough train/test splits and cross-validation to avoid overfitting.
Our contributions are:
\begin{itemize}
	\item We identify 682 \gt hosts, of which a large fraction exhibits variable clock skew
	\item We define novel features for sibling classification, capable of, \eg identifying and comparing variable clock skew
	\item We utilize thorough train/test methodology and machine-learning to build and evaluate classifier models
	\item We achieve excellent train and test performance even for hosts with variable clock skew 
	\item We establish scalability through large-scale measurements and find 149k server siblings
	\item We publicly share our \gt, code, and data
\end{itemize}
We structure this work as follows.
In Section\,\ref{sec:bgrel}, we discuss background and related work. 
We present our methodology in Section\,\ref{sec:meth}, and define features in Section\,\ref{sec:features}.
Section\,\ref{sec:models} presents and evaluates our models, followed by their large-scale application in Section\,\ref{sec:ls}.
We discuss outliers and influencing factors in Section\,\ref{sec:discussion}, concluded by Section\,\ref{sec:concl}.
\section{Related Work}\label{sec:bgrel}
{\noindent}We introduce background and related work in four categories: remote clock skew estimation, remote identification, IPv6-IPv4 comparative studies, and IPv6-IPv4 sibling detection.

\textbf{Remote Clock Skew Estimation: }%
Accurate time-keeping on computing machinery is a notoriously difficult problem: precisely oscillating hardware is prohibitively expensive for most machines. 
The dominant protocol to synchronize low-precision machines, NTP, exhibits many difficulties even after decades of development \cite{maloneleap,veitch2016ntp}.
Hence, clocks in most devices in the Internet do not run in sync with true time, but deviate from it to an extent that is measurable over the Internet. 
This deviation is called \textit{skew}, and can either be consistent over time (\textit{constant skew}), or vary over time (\textit{variable skew}).
Protocols or protocol extensions that include timestamps from a remote machine allow for measuring clock skew with good accuracy by comparing local and remote timestamps over time.
This skew can be used to remotely identify network devices.
Foundations in this field were laid by Paxson \cite{paxson98} in 1998 and Moon et al. \cite{moon1999estimation} in 1999.
Kohno et al.\,\cite{kohno2005remote} in 2005 first apply these techniques to TCP timestamps. 
They conduct a variety of case studies on the influence of external factors on timestamp behavior, e.g., power-saving or virtualization settings. 
Murdoch\,\cite{murdoch2006hot}, and Zander and Murdoch \cite{zander2008improved} actively induce variable skew on remote devices to identify Tor hidden services. 
However, their method of decision taking is tailored for few hosts and human inspection.

\textbf{Other Remote Identification Techniques: }
Determining whether a set of IP addresses belong to the same router is an important and well-understood problem in Internet research. 
Scientific tools such as Ally\,\cite{spring2002measuring}, Radar Gun\,\cite{radargun2008} and MIDAR\,\cite{midar2013} use IP Identification (\textit{IP ID}) header values to answer this question, exploiting the fact that the \textit{IP ID} counter is commonly shared between interfaces.
Unlike IPv4, IPv6 only offers \textit{IP ID} values in an extension header for fragmented packets (cf.\,RFC\,2460).
In 2013, Luckie et al. \cite{luckie2013speedtrap} publish \textit{speedtrap}, which uses forced packet fragmentation for alias resolution in IPv6.
In 2015, Beverly et al. \cite{ipv6router} use IPv6 identification values to measure router uptime. 

\textbf{IPv6-IPv4 Comparative Studies: }%
In 2015, Bajbai and Schönwälder\,\cite{bajpai2015ipv4} compare connection setup latency of domains for IPv6 and IPv4 addresses. They cite content being served from different machines as one of the potential reasons for latency differences.
In 2016, Czyz et al. \cite{czyzbackdoor} compare security characteristics at service level, \ie \textit{AAAA} and \textit{A} records of a domain. They find IPv6 addresses to frequently have worse security characteristics.

\textbf{IPv6-IPv4 Sibling Detection: }%
The problem of classifying sibling relationships at a machine level has first been tackled by Berger et al.\,\cite{berger2013internet}. 
Using customized DNS replies, they associate DNS client resolvers through opportunistic passive probing and open DNS resolvers through active probing. 
This technique only works on DNS clients or open resolvers, and requires a DNS server backend infrastructure.
In 2015, Beverly and Berger\,\cite{beverly2015server} refine prior work on remote clock skew estimation through TCP timestamps and apply it to actively probe IPv6-IPv4 servers for sibling classification. 
Their algorithm is as follows:
First, they filter non-siblings based on different TCP option signatures.
Second, they classify the kind of TCP timestamp behavior (e.g., random, monotonic, non-monotonic). 
Third, they compare the angle of two constant clock skews to determine a sibling/non-sibling relationship. 
They achieve very good metrics (99.6\% precision) on their training data, but acknowledge their comparably small ground truth \ds of 61 hosts might be prone to overfitting. 

They highlight the existence of hosts with variable clock skew, for which we provide precise classification features and models in this work. 
\section{Methodology}\label{sec:meth}
We put great care into avoiding overfitting and providing a sibling classifier that generalizes well.
First, we collect a sufficiently large and diverse \gt, significantly exceeding that of prior work. 
We then conduct a series of traffic measurements against this \gt and a \ls \ds. 
Next, we define features potentially suited to discern siblings and non-siblings. 
Subsequently, we develop sibling decision algorithms based on these features, leveraging both manual analysis and machine learning algorithms.
We train and evaluate those algorithms based on a train/test split in 10-fold cross-validation.

\textbf{Acquiring a \GT \DS: }\label{sec:gt}
A critical success factor of this work is to obtain a \gt \ds with numerous and diverse hosts.
As prior work does not publish their \gt \ds, we set out to construct our own by (i) collecting \gt servers from personally known operators and (ii) leveraging public frameworks which enforce an IPv6 and IPv4 address to reside on the same machine.
For the latter, we include \textit{RIPE Atlas} anchors\,\cite{ripeatlas} and \textit{NLNOG RING} probes\,\cite{nlnogring}.
Table \ref{tab:gt} lists our \gt data and compares it to related work.

Ripe Atlas anchors exist in two different hardware versions\,\cite{bajpai2015lessons}, which we split out as \textit{RAv1} and \textit{RAv2}.
Within the groups of \textit{RAv1} and \textit{RAv2}, there is no hardware or software diversity.
\textit{NLNOG Ring} (``\textit{ring''}) nodes are formed by installing a provided image onto a virtual or physical machine. 
The \textit{ring} group hence offers hardware diversity, but no software diversity.
Our \textit{servers} group offers soft- and hardware diversity.
The \textit{RAv1}, \textit{RAv2} and \textit{ring} groups offer good geographical diversity, while the \textit{servers} group centers on  Germany.

Please note that this \ds allows for testing both sibling and non-sibling relationships, as non-siblings can be created by mixing addresses from different servers. 

\begin{table}
	\caption{Our \gt \ds covers diverse Autonomous Systems (ASes), Countries (CC), and clock skew characteristics. Subsets can have hardware and/or software diversity.}
	\label{tab:gt}
	\centering
	\resizebox{\columnwidth}{!}
	{\begin{tabular}{lrrrrr}  
			\toprule
			\DS & Hosts& \#AS & \#CC & Skew & Div. \\
			\midrule
			2016-03 (``\textit{03}'') & 458&373&40 & variable & sw+hw\\
			\midrule
			2016-12 (``\textit{12}'')& 682 & 536 & 80 & variable & sw+hw\\
			~~\textit{servers} & 31 &9& 5& variable & sw+hw\\
			~~\textit{ring} & 430 & 383& 56  & variable & hw\\
			~~\textit{RAv1} & 12 & 12 & 11 & variable & -\\
			~~\textit{RAv2} & 209 & 192 & 64 & constant & - \\			
			\midrule			
			Beverly\,\cite{beverly2015server}& 61 & 34 & 19 & constant & unkn.\\
			\bottomrule
	\end{tabular}}%
\end{table} 

\textbf{Measurement Methodology: }%
To obtain a sufficient amount of fingerprints, we repeatedly connect to every sibling candidate IP address in parallel for a duration of ten hours, with the goal of acquiring at least one TCP Timestamp per minute. 
We open a HTTP connection and issue a \textit{GET research\_scan} query.
As this resource-heavy approach repeatedly requires a full TCP and HTTP connection for every IP address, we process batches of 10k IP address pairs for our large-scale measurements.
Our ground-truth measurements fit into one batch. 
As our methodology aims to identify similarities in clock skew, measurements to all IP address of a sibling candidate need to be conducted in the same batch.
We leverage the TCP keepalive option to avoid establishing a new connection every minute, but found many servers to quickly close our connections after few keepalive packets.

Our measurement stack consists of a Python3 master that dispatches work to several processes, which in turn start one \textit{{urllib3}} thread per target IP address.
Moving to a C library or high-speed packet-processing frameworks such as DPDK \cite{dpdk} or libmoon\,\cite{moongenimc} might significantly reduce the kernel packet processing overhead and allow for larger batches.
We acknowledge that our measurements might be considered intrusive, which we discuss later in this  Section.

\textbf{Measurement Runs: }%
In March 2016, we conduct one run against our 2016-03 \gt, referred to as \textit{gt1}.
In December 2016 and January 2017, we conduct six measurements against the 2016-12 ground truth, referred to as \textit{gt2} through \textit{gt6}.
Notable are the runs \textit{gt4} through \textit{gt6}, which cover the time before, during and after the leap second on December 31, 2016. 
We also conduct a \ls measurement campaign in August 2016, which we further discuss in Section \ref{sec:ls}.
As the measured offset is relative to the offset of our own clock, we usually disable \textit{ntpd} to avoid creating \nl offsets from clock adjustments on our machine.
We enable \textit{ntpd} during the \textit{gt2} measurement to test this hypothesis.

\textbf{Training and Evaluation Methodology: }%
Training an algorithm on a few hundred \gt hosts that will later be applied on millions of hosts comes with two challenges:
First, the ground truth obtained might not be a representative sample of the full population of hosts in the Internet. 
Second, classifier training may overfit the training data, achieving very good train/test metrics on the ground truth, but failing on \ls application.
We aim to mitigate the risk of training on a non-representative sample by establishing software, hardware, geographical, and administrative diversity in our \gt. 
We find our \gt to exhibit diverse TCP timestamping characteristics even when compared to our \ls \ds. 
To avoid overfitting our \gt for both machine-learned and manually assembled decision algorithms, we deploy a strategy of train/test splits and cross-validation. 
We also aim to minimize model complexity to provide better generalization.%

\textbf{Ethical Considerations: }%
\noindent We follow an internal multi-party approval process, among others based on Partridge and Allman\,\cite{partridge2016ethical}, before any measurement activities are carried out. 
We conclude that our measurements and the resulting data can not harm individuals, but may result in investigative effort for system administrators.
We aim to minimize this effort by deploying scanning best-practice efforts of (i) using dedicated scan machines with explanatory websites, (ii) maintaining a blacklist, (iii) reply to every abuse e-mail (seven received, one asking for blacklisting, six curious), and (iv) request URLs preceded by \textit{{/research\_scan}} to allow quick identification of our connections.
Furthermore, based on user discussions, we will respect \texttt{\footnotesize{robots.txt}} and set a descriptive HTTP user agent in future work. 
To conclude, we argue that no individual was harmed by our active measurements. 
We also conclude that the gathered data bears little privacy intrusion, and hence release all data that was based on publicly available sources.
%
\section{Features}\label{sec:features}
In this section, we present the set of features investigated and later leveraged by our algorithms for sibling detection. 
We present the features in the order they are calculated, as some features depend on the existence of others. 

It is important to note that there is a distinction in the nature of these features. 
Namely, a feature can be either \textit{falsifying} or \textit{verifying}. 
\textit{Falsifying} features may only help to determine a non-sibling relationship, whereas \textit{verifying} features can actually determine a sibling relationship with confidence.

For every group of features, we explain their calculation and list their specific outputs, where \textit{output$_6$} indicates an IPv6 feature, and \textit{output$_4$} an IPv4 feature.

\textbf{Network Level Features: }%
We test various network level features, such as network latency, initial Time-to-Live values, OS predictions, or open ports. We find those features to have very low discriminative power and exclude them from further analysis. Our technical report gives details on these measurements and the obtained results\,\cite{rouhi16}.

\textbf{TCP Options Fingerprint: }%
Similar to Beverly and Berger\,\cite{beverly2015server}, we leverage TCP options as a first falsifying feature. 
We compare the \emph{presence} and \emph{order} of options and the \textit{no operation (NOP)} padding bytes. 
Additionally, if the \textit{window scale} option is present, we consider its value for the process of falsifying non-siblings, as it has demonstrated high discriminative capability in our test \ds.
We find values of some options such as \textit{MSS} to frequently differ by non-static offsets even for ground truth siblings. Thus, we do not include those in this filtering step.
As an example, we frequently find the option fingerprint \texttt{MSS-SACK-TS-NOP-WS07} in our \gt \ds, and the \texttt{MSS-NOP-WS08-SACK-TS} in our \ls \ds.
We highlight that asking for more exotic options slightly increases diversity in answers, but we did not find the effect strong enough to justify the additional measurement overhead.
\textbf{Features}: \textit{Options$_4$}, \textit{Options$_6$},  $\textit{opts\_diff} = !(\textit{Options$_4$} == \textit{Options$_6$)}$.

\textbf{Remote TCP Timestamp Clock Frequency: }%
In a next step, we calculate the remote clock frequency as employed by Kohno et al.\,\cite{kohno2005remote}.
We first calculate relative remote TCP timestamps (32-bit unsigned integers) as $v_i=T_i-T_1$, where $T_i$ is the TCP timestamp contained in the $i$-th received packet.
Then, the relative received timestamps are calculated as $x_i = t_i - t_1$, where $t_i$ is the time the $i$-th packet was observed by the prober and $x_i$ is in seconds.
We check the resulting array $[x, v]$ for monotonicity and fix wrap-arounds. 
We then solve a linear regression against $[x ,v]$, where the resulting slope is the remote clock frequency \textit{Hz}. 
We find typical values of 10, 100, 250 and 1000 \textit{Hz}, all within the  range of 1 to 1000 Hz as in RFC\,7323. 
We also save the $R^2$ value of the linear regression, $R_\text{\textit{Hz}}^2$. Low or different $R_\text{\textit{Hz}}^2$ values may indicate erratic time-stamping behavior such as randomized timestamps. 
We expect a sibling's clock to tick with the same frequency for IPv6 and IPv4.
Hence, for each sibling candidate, we calculate the difference between $\text{\textit{Hz}}_4$ and $\text{\textit{Hz}}_6$ as a falsifying metric.
This produces one false negative occurrence in the ground truth \ds, which we discuss in Section \ref{sec:discussion}.
\textbf{Features}: $\text{\textit{Hz}}_4$, $\text{\textit{Hz}}_6$, \textit{hz\_diff}, $R_\text{\textit{Hz4}}^2$, $R_\text{\textit{Hz6}}^2$.

\textbf{Raw TCP Timestamp Value: }%
As a next step, we compare the raw TCP timestamp values $T_1^4$ and $T_1^6$ of a sibling candidate pair. 
As the TCP timestamp clock is, except for wrap-arounds, typically monotonically increasing, the raw value of the 32-bit timestamp offers a certain level of entropy across hosts. We expect the values for a sibling to be very close for IPv6 and IPv4 as they are generated from the same clock.
Using the \textit{Hz} values calculated in the previous paragraph, we can calculate the absolute difference between two remote clocks using Equation \ref{eq:tcpraw}:
\begin{align*}
\centering
\Delta_{tcp}  &= T_1^4 / \text{\textit{Hz}}_4 - T_1^6 / \text{\textit{Hz}}_{6} &[s] \\
\Delta_{rec} &=  t_i^4  - t_i^6   &[s] \numberthis \label{eq:tcpraw}  \\
\Delta_{tcpraw}  &= | \Delta_{tcp} - \Delta_{rec}|  &[s] 
%
\end{align*}
First, we convert the raw TCP timestamps to seconds by dividing through \textit{Hz}, 
Second, we calculate the difference between the local received timestamps. 
The final metric $\Delta_{tcpraw}$ is obtained by computing the absolute difference between the two values mentioned above. 
This metric can be interpreted as the time difference between the last TCP timestamp counter reset for IPv6 and IPv4. 
\textbf{Feature}: $\Delta_{tcpraw}$.

\textbf{Clock Offset and Skew Calculation: }%
In a next step, we can estimate the remote clock offset, \ie the deviation of a remote clock from its expected values.
To do so, we calculate the expected relative remote time $w_i = v_i/\text{\textit{Hz}}$ and the offset to the observed time $y_i = w_i - x_i$, both measured for the $i$-th observed packet. 
The resulting array $[x_i, y_i]$ is then used for more fine-grained investigations such as clock skew estimation, which will be explained in the following paragraphs.  
This array is also used for plots in this work. 

As noted by Kohno et al \cite{kohno2005remote}, the derivative of this array would theoretically be the \textit{skew} of the remote clock. 
However, due to delay variances and various other effects, it is not sound to form the derivative of this array, but we rather use a more stable method to obtain the skew.

In this work, we use Robust Linear Regression~\cite{theil1992rank} to obtain a robust and outlier-resistant regression, whose slope $\alpha$ we use as an estimation of remote clock skew. 
The rationale behind using this method is that offset points are in nature heavily impacted by various network dynamics \cite{paxson98} and hence prone to outliers.
Additionally, we store $R^2_{skew}$ which is the linear regression's coefficient of determination and is used to estimate the quality of the line fitted by the regression.
\textbf{Features}: $\alpha_4$, $\alpha_6$, $\alpha_\textit{diff}$, $R^2_\textit{skew4}$, $R^2_\textit{skew6}$, $R^2_\textit{skewdiff}$.

\begin{figure}[h]
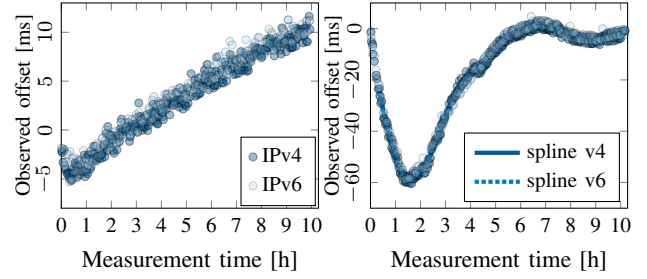

	\begin{subfigure}[b]{0.45\columnwidth}
		\input{figures/tcptimestamp/lin_skew1.tex}
		\vspace{-5mm}
	\end{subfigure} 
	\begin{subfigure}[b]{0.45\columnwidth}
		\input{figures/tcptimestamp/spline_lrg_dynam1.tex}  
		\vspace{-5mm}		
	\end{subfigure}
	\caption{Siblings with constant skew and small dynamics (left), and variable skew and large dynamics (right).}
	\label{fig:skewconstvar}
\end{figure}

\textbf{Calculation of Dynamic Range: }%
Another feature we consider is the \textit{dynamic range} of the offset array: 
While some hosts exhibit an offset of several seconds over the course of 10 hours, other hosts exhibit an offset of few milliseconds (cf. Figure \ref{fig:skewconstvar}). 
As this dynamic might be valuable information, we aim to extract it as a feature. 
To calculate this dynamic range in a manner that is stable against latency-caused outliers, we first prune the top and bottom 2.5\% of offset arrays, and then store the difference between maximum and minimum as \textit{rng}$_4$ and \textit{rng}$_6$.
\textbf{Features}: \textit{rng}$_4$, \textit{rng}$_6$, 
$\textit{rng\_diff}$=$|\textit{rng}_4$-$\textit{rng}_6|$, 
$\textit{rng\_avg}$=$(\textit{rng}_4 + \textit{rng}_6)/2$, 
 $\textit{rng\_diff\_rel}$=$\textit{rng\_diff}/\textit{rng\_avg}$.

\textbf{Variable Skew Calculation: }
While $\alpha$ and $R^2_{skew}$ fuel sibling/non-sibling classification for hosts with constant skew, additional steps are necessary to gain insight into the behavior of sibling candidates with variable skew.

\begin{figure}[H]
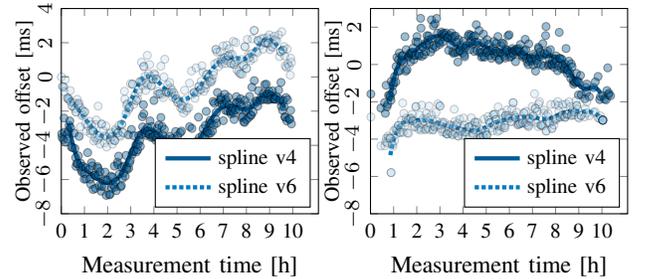

	\centering
 	\begin{subfigure}[b]{0.45\columnwidth}
 		\input{figures/tcptimestamp/splinefit_nomap.tex}
 		\label{fig:splinefit_nomap}
 	\end{subfigure}
 	\begin{subfigure}[b]{0.45\columnwidth}
 		\input{figures/tcptimestamp/splinefit_nomap_nonsib.tex}
 		\label{fig:splinefit_nomap_non-sib}
 	\end{subfigure}
 	\caption{Splines for sibling (left) and non-sibling (right).}
 	\label{fig:splinefit_sib}
\end{figure}

To approach variable skew, we fit a polynomial spline against the $[x_i, y_i]$ arrays of a sibling candidate.
Among various options to fit polynomial splines, we find it well-suited to pick 13 equidistant offset points and fit cubic splines between these candidate points in an approximative manner.
Figure~\ref{fig:splinefit_sib} shows the curve fitting approach for both siblings and non-siblings. 
In the next step, we minimize the area between the two splines by shifting the \textit{y}-offset of one slope. 
This minimal area between the v4-spline and the v6-spline, \textit{spl\_diff}, is an output of our variable skew calculation. 
As the area between the two splines is also proportional to the dynamics of offsets, we also provide scaled version $\textit{spl\_diff\_scaled}=\textit{spl\_diff}/\textit{rng\_diff}$.
\textbf{Features}: \textit{spl\_diff}, \textit{spl\_diff\_scaled}.
%
%
\section{Classifier Model Training and Evaluation}\label{sec:models}
Based on the discussed features, it is possible to train classifier \textit{models}, that can be used to predict whether a pair of IP addresses is a sibling or not. 
In this section, we explain our approach to train and evaluate various such models, preceded by an explanation of our evaluation methodology.
\subsection{Model Evaluation}%
To evaluate classifier models, a wide range of metrics exists, of which we focus on two:
First, we prioritize high \textit{precision}, defined as $tp/(tp+fp)$, \ie a low number of false positives. 
With higher precision, fewer non-siblings will be among predicted siblings. 
Second, we also want metrics for other aspects of performance to be stable to avoid overprioritizing precision.
For this, we use the Matthews Correlation Coefficient \textit{(MCC)}\,\cite{mcc-sklearn} as defined in Equation \ref{eq:mcc}.
This coefficient nicely and robustly factors in all kinds of aspects of classifier performance for binary problems, where 1 would be a perfect score and 0 the worst, \ie a coin flip.
\begin{equation}
\label{eq:mcc}
\text{MCC} = \frac{ tp \cdot  tn - fp \cdot fn } {\sqrt{ (tp + fp) ( tp + fn ) ( tn + fp ) ( tn + fn ) } }
\end{equation}

For training and evaluation purposes, we generate around 400k\footnote{$(n\cdot(n-1)), n=682$, we use both the v6-v4 and v4-v6 combination}  non-siblings from our \ds by mixing the IPv6 address of one sibling with the IPv4 addresses of other siblings. 
We generate the maximum possible number of non-siblings for each evaluation, and then equally weight both output classes for the classifier training. 
We generate non-siblings only for siblings actually used, \eg we first split siblings into train and test, and then form the non-siblings based on these splits. 

Table \ref{tab:gt-eval-hl} shows the evaluation results for our different models. 
Table \ref{tab:gt-eval-groups} investigates whether a model's results are dependent on the evaluated subgroup.

Having established our evaluation approach, we now discuss filtering steps applied during feature calculation, followed by the individual models.
\subsection{First-Order Filtering for All Models}%
\label{sec:firstorderfilters}
Certain sanity checks on first-order filters apply to all mod\-els, \eg if the calculation of \textit{Hz} fails, many dependent features can not be calculated.

\textbf{Different TCP Options}: If sibling candidates offer different TCP options, our algorithm stops with a non-sibling decision before continuing with feature calculation. 

\textbf{Hz Calculation}: When calculating the remote clock frequency, the linear regression applied to do so may fail for reasons such as randomized timestamps. To only incorporate sound remote Hz frequencies, we require the $R^2$ parameter to be above $0.9$. 
We classify sibling candidates with failed Hz calculations as non-siblings.

\textbf{Different Hz Values}: When a sibling candidate offers different \textit{Hz} values for IPv6 and IPv4, calculation of dependent features becomes meaningless. We hence decide candidates with different Hz values as non-siblings.

\textbf{Too Small Hz Values: } When a Hz value below 1\,Hz is calculated, we also stop further calculation and decide for a non-sibling relationship.

In all these cases, we decide for a non-sibling relationship and stop calculating other features. This works with a very low false negative rate (4 in 682) for our \gt \ds.
We discuss those outliers in Section \ref{sec:outliers}.

\subsection{Beverly/Berger Algorithm}

For comparison, we implement the sibling detection algorithm as proposed by Beverly and Berger~ \cite{beverly2015server}.
The algorithm is designed for constant skew and works primarily by comparing the constant skew of sibling candidates.

The algorithm does not find siblings with high precision ($<1$$\%$, see Table \ref{tab:gt-eval-hl}) in our combined \ds, which includes many hosts with variable clock drift.
We find the algorithm's performance to vary slightly for different groups of our test set (see Table \ref{tab:gt-eval-groups}). 
As this variation is not systematic (\ie due to overwhelming group characteristics), we argue that this is based on circumstantial existence of hosts that are well-fit to Beverly and Berger's algorithm. 

\begin{table}
	\caption{Hand-Tuned and Machine-Learned Classifiers train and test very well, speaking to good generalization. Beverly algorithm is not generalizing well to the more diverse \ds.}
	\label{tab:gt-eval-hl}
	\centering
	\resizebox{\columnwidth}{!}
	{\begin{tabular}{lrrrrr}  
			\toprule
			Algo. & Train DS & Test DS & Prec. & MCC & Type \\
			\midrule
			Bev. 			& Bev. 		 & Bev. & 99.6\% & n/a & Train \\
			Bev. 			& Bev. 		 & 03$\cup$12 & .9\% & .17 & Test \\  
			HT1 		& 03 & 03 & 100\% & .99 & Train\\			
			HT1 		& 03 & 12$\setminus$03 & 99.49\% & .98 & Test\\
			ML1 		& 03$\cup$12  & 03$\cup$12  & 99.36\%  & 1.0 & Train \\
			ML1 		&  03$\cup$12  & 03$\cup$12  & 99.88\% & 1.0 & Test	 \\
			\bottomrule
			\multicolumn{6}{p{\columnwidth}}{\footnotesize{Legend: HT1 denotes our hand-tuned algorithm. 03 denotes the 2016-03 \ds, 12 the 2016-12 \ds. ML1 values are the arithmetic mean of 10-fold cross-validation. All tests against 2016-12 are arithmetic means over the results from measurement  runs 2 through 7. Calculation of MCC for Bev. data set not possible from metrics given by Beverly and Berger~ \cite{beverly2015server}.}}\\
	\end{tabular}}%
\end{table} 

\begin{table}
	\caption{Performance of Beverly algorithm slightly dependent on group, our hand-tuned and machine-learned (not shown) algorithm independent from group.}
	\label{tab:gt-eval-groups}
	\centering
	\resizebox{\columnwidth}{!}
	{\begin{tabular}{lrrrrr}  
			\toprule
			Algo. & Train DS & Test DS & Prec. & MCC & Type \\
			\midrule
			Bev. 			& Bev. 		 & 03$\cup$12-server & 8.33\% & .17 & Test \\  
			Bev. 			& Bev. 		 & 03$\cup$12-ring & 1.09\% & .09 & Test \\  
			Bev. 			& Bev. 		 & 03$\cup$12-rav1 & 8.35\% & .00 & Test \\  									
			Bev. 			& Bev. 		 & 03$\cup$12-rav2 & .79\% & .05 & Test \\  
			\midrule
			HT1 		& 03 & 12$\setminus$03-server & 100\% & .99 & Test\\
			HT1 		& 03 & 12$\setminus$03-rav2 & 99.16\% & .98 & Test\\
			HT1 		& 03 & 12$\setminus$03-ring & 100\% & .98 & Test\\
			HT1 		& 03 & 12$\setminus$03-rav1 & 100\% & 1.0 & Test\\									
			\bottomrule
	\end{tabular}}%
\end{table}

\subsection{Hand-Tuned Decision Algorithm}
One of our classifiers is a hand-tuned decision algorithm similar to Beverly and Berger's\,\cite{beverly2015server}.
We hand-tune our algorithm against our \textit{2016-03} \gt and test it against the newly added hosts of the \textit{2016-12} \gt. 
This results in a 40\% train and 60\% test set, with all subgroups achieving $>$40\% test size. 

\begin{algorithm}
	\caption{Hand-Tuned Sibling Classification Algorithm (trained and tested on disjunct \dss).}
  	\begin{algorithmic}[1]
  		\State \textbf{if} $\textit{opts\_diff} \implies$ non-sibling. 
  		\State \textbf{if} $\textit{Hz}$ calculation failed or invalid $\implies$ non-sibling.
  		\State \textbf{if} $\Delta_{tcpraw} > z_1 \implies$ non-sibling 
  		\State \#\#\# Linear Testing (using Robust Linear Regression \cite{theil1992rank})
  		\State \textbf{if }$R^2_{skew4} \geq z_2 \land R^2_{skew6} \geq z_2$:  
  		\State \indent \textbf{if} $\textit{sign}(\alpha_4) \neq $ $\textit{sign}(\alpha_6) \implies$ non-sibling. 
  		\State \indent \textbf{if} $|\alpha_\textit{diff}| \leq z_4 \implies$ sibling. 
  		\State \textbf{else if} $R^2_{skew4}  \geq  z_2 \oplus R^2_{skew6} \geq  z_2$: 
  		\State \indent \textbf{if} $|R^2_\textit{skewdiff}| \geq z_3 \implies$ non-sibling. 
		\State \#\#\# \NL Testing:
  		\State \textbf{if} $\textit{rng}_4 \leq z_5 \land \textit{rng}_6 \leq z_5 \implies$ unknown. 
  		\State \textbf{if} $\textit{rng}_4 \geq z_5 \oplus \textit{rng}_6 \geq z_5$: 
  		\State \indent \textbf{if} $\textit{rng\_diff} \geq z_6 \implies$ non-sibling. 
  		\State \textbf{if} $\textit{rng}_4 \geq z_7  \land \textit{rng}_6 \geq z_7$: 
  		\State \indent \textbf{if} $\textit{spl\_diff} \leq y_1 \implies$ sibling. 
  		\State \indent \textbf{else} $\implies$ non-sibling.
  		\State \textbf{if} $\textit{spl\_diff} \leq y_2 \implies$ sibling. 
  		\State \textbf{if} $\textit{spl\_diff} > y_3 \implies$ non-sibling. 
  		\State \textbf{else}  $\implies$ unknown.
  	\end{algorithmic}
  \hrule\vspace{1mm}
Values used: $z_1=1$, $z_2=.81$, $z_3=.2$, $z_4=.00005$,  
$z_5=1.5$, $z_6=.47$, $z_7=14$, $y_1=2.3, y_2=.6, y_3=4.0$
\label{algo:decision algorithm}
\end{algorithm}%
The formalized algorithm is displayed in Algorithm~\ref{algo:decision algorithm}.
The following provides a terse description of its high-level decision taking steps.
The algorithm offers many subtleties, and we recommend our source code and tech report \cite{rouhi16} as a detailed reference. 
Similar to Beverly and Berger, we first eliminate candidates with different TCP options.
Then, our algorithm performs first-order filtering a described in Section \ref{sec:firstorderfilters}.
Third, we eliminate candidates with raw TCP timestamps too far apart.
In line 5, we test whether to apply linear logic by evaluating the $R^2_\textit{skew}$ values of robust linear regression. 
Skews with differing slope signs are classified as non-siblings (line 6), whereas small slope differences are classified as siblings (line 7).
In line 8 and 9 we classify those cases as non-siblings if only one skew is clearly constant and there is a large difference in $R^2_\textit{skew}$ values.

If linear testing was not conducted or not decisive, we apply nonlinear testing. 
For this, we first test the dynamics of both signals to exclude cases with negligible (line 11) or very different dynamics (lines 12 and 13). 

We take further decision based on the minimal area between non-linear splines in lines 14 to 18. 
Based on whether the overall dynamics are large (line 14-16) or small (line 17-19), we apply different thresholds. 
We found this simplistic distinction between large and normal dynamics to provide good results on our \ds, but acknowledge that this step could potentially be improved by means of finer tuning, for example by scaling the threshold by the dynamics. 
Our algorithm, similar to Beverly's and Berger's, features some guard intervals. 
In those, we can not take a meaningful sibling/non-sibling decision and decide for unknown.

As visible in Tables \ref{tab:gt-eval-hl} and \ref{tab:gt-eval-groups}, our algorithm achieves very good ($>$$99\%$ precision, $\geq$$.98$ MCC) metrics in training and testing and is insensitive to subgroups. 
We argue that this algorithm likely generalizes well to new \dss.

\subsection{Decision Tree}%
Using scikit-learn~\cite{sklearn}, we train a CART Decision Tree on our features described in Section~\ref{sec:features}.
For each of the seven measurement runs \textit{gt1} through \textit{gt7}, we do a 10-fold cross-validation with proportional selection from all subgroups (\textit{servers}, \textit{ring}, \textit{rav1}, \textit{rav2}).
We find all models to consistently perform well with low variance, and report the arithmetic mean across validation folds and measurement runs for Table \ref{tab:gt-eval-hl}. We also find all models' performance to be independent of groups.

For model selection, we export all generated decision trees and find very little variance:
All trees contain just one branch, where they use a single threshold against the $\Delta_{tcpraw}$ feature to decide for sibling or non-sibling, \ie as a \textit{verifying} metric.
Our hand-tuned algorithm used a threshold of $>$$1s$ as a \textit{falsifying} metric.
We find the majority of models to pick a threshold of $>$$0.2557s$ for non-siblings and pick that value for our final model. Please note that the \textit{ML1} model is preceded by the first-order filters described in Section \ref{sec:firstorderfilters}.

We argue that this model, due to its simplicity, will likely generalize best and recommend it for further use.
\section{Large-Scale Application \& Results}\label{sec:ls}
We apply our measurement methodology and classifier models to a \ls \ds to evaluate their scalability and suitability for finding sibling pairs for large-scale structural Internet studies. 
We first identify sibling candidates by resolving \textit{A} and \textit{AAAA} records for 162M domains, obtained from both registrars and ``drop list'' resellers. %
We filter blacklisted IP addresses, and form sibling candidates from all possible \textit{A} and \textit{AAAA} combinations per domain. 
Table \ref{tab:ls1} details the statistics of this process by top-level domain. 
\begin{table}
		\caption{Statistics of Large-Scale Domain Scans.}
		\resizebox{\columnwidth}{!}
			{\begin{tabular}{lrrrr}
    	\toprule
		Source & \#Domains & \#IPv4 RRs & \#IPv6 RRs  & \#Candidates\\
		\midrule
		Alexa  &  1M & 1.2M & 108k & 191k \\
		biz  &  2.2M & 2.0M & 82k & 104k \\
		com  &  127M & 134M & 4.4M & 6.7M \\
		info  &  5.5M & 5.0M & 270k & 299k \\
		mobi  &  682k & 560k & 12k & 15k \\
		net  &  15.7M & 14.3M & 630k & 898k \\
		org  &  10.8M & 10.6M & 464k & 700k \\
		xxx  &  101k & 178k & 611 & 892 \\
		\midrule
		Total & 162.4M & 167.8M & 5.9M & 8.9M \\
		\bottomrule
		\label{tab:ls1}
		\end{tabular}}%
\end{table}

We find the number of sibling candidates to be bound by the relatively few \textit{AAAA} records:
We obtain only 6M \textit{AAAA} records, compared to 168M \textit{A} records. 
The number of sibling candidates, as the cross-product of \textit{A} and \textit{AAAA} records, will multiply with increased IPv6 deployment.
Processing of the resulting 8.9M candidates is quantified in Table \ref{tab:ls2}.
\begin{table}
\caption{Statistics of Large-Scale Domain Scans.}
{\begin{tabular}{lrr}
\toprule%
\label{tab:ls2}%
Processing Step & IPv4 & IPv6\\
\midrule
Sibling candidates & \multicolumn{2}{c}{ 8,893,132 } \\
Unique candidates &  241,085 & 372,607 \\
tcp80 responsive &  226,419 & 315,782 \\
Timestamp-capable &  128,420 & 216,350 \\
Consistent TCP options &  128,146 & 216,073 \\
Remaining candidates & \multicolumn{2}{c}{ 6,619,100 } \\
Unique candidate pairs &\multicolumn{2}{c}{ 371,071 } \\%
Unique addresses in pairs &  95,469 & 212,700 \\
Pairs with measurement results &\multicolumn{2}{c}{ 351,994 }\\
\bottomrule%
\end{tabular}}%
\end{table}
\\
After processing our sibling candidates to unique IPv6 and IPv4 addresses, we scan those addresses with zmap \cite{durumeric2013zmap} on port TCP/80. 
We leverage our previously developed IPv6-capable version of zmap for this \cite{Gasser2016ipv6}.
We find the majority of IP addresses to be responsive on port TCP/80. Discovery on more ports, such as TCP/443, might yield more responsive hosts.
We next eliminate machines that do not offer the TCP Timestamp option, which is a prerequisite to applying our technique. 
We find 57\% (IPv4) and 69\% (IPv6) of responsive IP addresses to offer the TCP Timestamp option.
The higher percentage for IPv6 could be caused by IPv6 being offered by newer machines with more modern TCP configurations.

As we use the TCP option fingerprint of a remote host to filter for non-siblings, we extend zmap with TCP options capabilities. 
We chose to form a complex TCP options payload as this offers more possibilities for different TCP stacks to offer different replies. 
We ask for the set of options of: \texttt{$<$SACK permitted,} \texttt{Timestamps,} \texttt{Window Scale,} \texttt{TCP Fast Open, Unknown, MPTCP$>$}. 
We include an unknown TCP option (by using a reserved option identifier) as this may also trigger a range of different responses, from simple mirroring to correctly dropping the unknown option.
However, this step removes only few hundred non-siblings in this \ls \ds. 

We reassemble sibling candidates with both usable IPv6 and IPv4 addresses, resulting in 6.6M sibling candidates. 
Those 6.6M sibling candidates represent 371k unique IPv6-IPv4 address combinations. 

In the next step, we measure the 371k unique candidate pairs by dividing them into batches of 10k addresses.
Through these measurements, we obtain a sufficient count of timestamps for 351k of 371k address pairs.
\begin{table}[tb]
		\caption{Large-Scale Domain Scan ($n=351,994$)}
		{\begin{tabular}{lrrr}
				\toprule
				Decision & HT & Bev. & ML1 \\
				\midrule
				Siblings & 57k & 203k &  149k\\
				Non-Siblings & 126k & 4k & 57k \\
				Unknown/Error & 169k & 145k& 143k \\
				\bottomrule
			\end{tabular}}
			\label{tab:ls3}
\end{table}

We then apply all discussed algorithms on this \ds and display the results in Table \ref{tab:ls3}.
Several conclusions stem from its analysis:
First, our \textit{HT} algorithm was possibly tuned too conservatively and only identifies about a third of the siblings identified by our \textit{ML1} algorithm.
Second, our reproduction of Beverly's and Berger's algorithm produces the most siblings, but its very low precision numbers as evaluated before likely cause these to be mainly false positives. 
Third, the amount of unknown/error decisions is surprisingly high. 
In the ground truth evaluation, we mapped those decisions to a non-sibling decision to allow for binary evaluation.
As we can not evaluate the large-scale measurement against a ground truth, we display the unknown/error category.
Future work might dig into these unknown/error cases, try to find a \gt, and perform further optimization. 
We investigate the contrast of siblings for \textit{HT} and \textit{ML1}, and find 57k siblings to intersect. 
Only few hundred of the \textit{HT} siblings do not intersect, caused by the more aggressive $\Delta_{tcpraw}$ threshold in \textit{ML1}.

Coming back to our initial goal, finding a confident set of siblings to study Internet-wide structural behavior, we conclude that both the \textit{ML1} and the \textit{HT} can find a significant number of siblings in Internet-wide scans.
For model selection, we repeat our argument that the simplicity of the \textit{ML1} makes it likely to generalize best, and the \textit{HT} model likely suffers from a high false negative rate on our \ls \ds.
%
\section{Outliers \& Discussion}\label{sec:discussion}
In this section, we analyze outliers in our \gt and discuss influence of several factors onto our measurements. 

\subsection{Analysis of Ground Truth Outliers}\label{sec:outliers}
We discuss the few outliers in our ground truth measurements and evaluation. 
First, the Ripe Atlas node \#6220 across measurements (i) returns different Hz values for IPv6 and IPv4 (100 and 1000), and (ii) has a significant ($\sim$$2^{28}$) raw TCP timestamps difference, equaling  $\sim$$3$ days difference at 1000 Hz). We have contacted the operators of this node to possibly obtain an explanation for this behavior.

Second, we find the hosts \textit{ovh0X.ring.nlnog.net} to return varying TCP options for IPv4 addresses through a measurement, typically varying between only \texttt{MSS} and \texttt{MSS-SACK-TS-NOP-WS07}.
Using tracebox\,\cite{detal2013revealing}, we typically receive responses that strip all but the MSS option at the last or penultimate hop. 
We conduct traceroutes path measurements from these machines and find the default IPv4 route traversing several RFC\,1918 IP addresses, possibly indicating a NAT or tunnel techniques interfering with our measurements. 
It is unclear why the hosts sometimes proceeds with a full set of TCP options.

We consider both hosts legitimate cases for our classifiers to take a non-sibling decision. 
While Ripe Atlas and NLNOG Ring ensure that the associated IP addresses reside on the host, deployed middleboxes or proxies seem to distort this sibling relationship. 
Hence, we consider it positive that our models did not classify these cases as siblings. 
 \begin{figure*}[!tb]
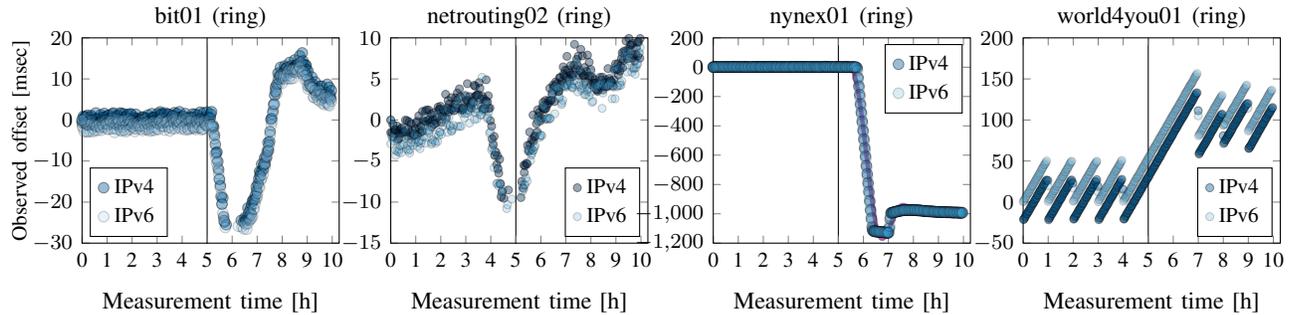

	\centering
	\begin{subfigure}[b]{0.225\textwidth}
		\input{figures/leapseconds/plots_tex.bit01.ring.nlnog.net-9065417606542677966.tex}
	\end{subfigure}%
	\begin{subfigure}[b]{0.225\textwidth}
		\input{figures/leapseconds/plots_tex.netrouting02.ring.nlnog.net--6616766576734116169.tex}
	\end{subfigure} 
	\begin{subfigure}[b]{0.225\textwidth}
		\input{figures/leapseconds/plots_tex.nynex01.ring.nlnog.net-3974175742968011748.tex}  
	\end{subfigure} 
	\begin{subfigure}[b]{0.225\textwidth}
		\input{figures/leapseconds/plots_tex.world4you01.ring.nlnog.net--686520199934996791.tex}
	\end{subfigure}
	\vspace{-.5cm}
	\caption{Leap second observations of four NLNOG ring siblings.}
	\label{fig:leapseconds}
\end{figure*}
\subsection{Influence of Measurement Machine's Clock Skew}
Irregularities in the measurement machine's clock may influence our measurements. 
We aim to minimize those irregularities by maintaining thermal conditions for the period of a measurement and by disabling the \textit{ntpd} daemon during our measurements. 
We conduct one measurement run (\textit{gt2}) with \textit{ntpd} enabled and find \textit{ntpd} interventions to be visible in manual analysis (through \nl dynamics replicated across all hosts). 
However, all classifier models returned equally good results for this measurement run.

\subsection{Influence of Leap Seconds}
We conduct measurement runs before, during, and after the leap second on Dec 31, 2016.  We find the metrics of our classifiers to be invariant to this circumstance, but interesting effects appear from visual inspection.
Figure \ref{fig:leapseconds} shows clock offsets for the measurement during the leap second, which happens 5 hours into the measurement and is marked by a vertical line.
We show four hosts with interesting behavior, most ($>$$99\%$) servers show no effect from the leap second at all. This is expected behavior, as the TCP timestamping clock is supposed to monotonically tick without regard to leap seconds.
Host \textit{bit01} shows a typical reaction unaware of leap seconds, with \textit{ntpd} adjusting clock speed after the leap second. 
Host \textit{netrouting02} seems to smooth out the leap second by starting to slow its clock about an hour before the leap second, a technique similar to the \textit{leap smear} deployed by Google~\cite{googleleapsmear}.
Host \textit{nynex01} reacts to the leap second with some delay, probably caused by periodic \textit{ntpd} polling. Remarkably, it rapidly adjusts its clock by a full second with some minor corrections following.
Host \textit{world4you} periodically adjusts its clock by a hard change instead of changing the tick speed. 
For some time after the leap second, no clock change is conducted, likely until local time has surpassed its remote equivalent. 

\subsection{Influence of Ripe Atlas Hardware Version}
\begin{figure}
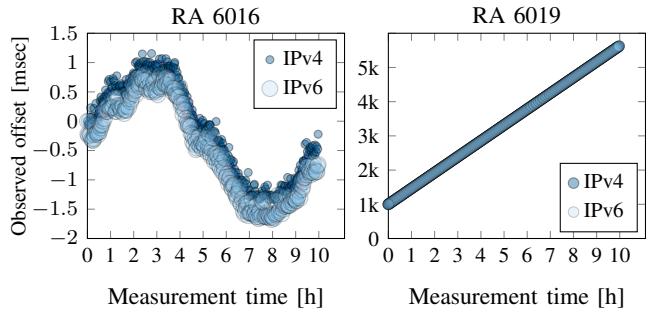

	\begin{subfigure}[b]{.45\columnwidth}
		\input{figures/ra/plots_tex.RA_6016-752441353800632200.tex}
	\end{subfigure} 
	\begin{subfigure}[b]{.45\columnwidth}
		\input{figures/ra/plots_tex.RA_6019--6905216013479891187.tex}
	\end{subfigure}
\vspace{-.5cm}
\caption{Ripe Atlas nodes with variable skew for \textit{v1} (left) and constant skew for \textit{v2} (right).}
\label{fig:ra_skew}
\end{figure}

Ripe Atlas anchors exist in two hardware versions which offer different characteristics\,\cite{bajpai2015lessons}.
Interestingly, remotely measuring the TCP timestamps of Ripe Atlas anchors reveals their hardware version, as all \textit{v1} anchors exhibit variable skew, while all \textit{v2} anchors offer constant skew (cf. Figure \ref{fig:ra_skew}).
\section{Conclusion and Future Work}\label{sec:concl}
We systematically approach the classification of IPv6-IPv4 servers siblings through active measurements, mainly reliant on TCP timestamping to estimate remote clock skew.
We extract a variety of features from our active measurements and feed these into (i) a reproduction of existing work's algorithm, (ii) a hand-tuned algorithm developed by us, and (iii) machine-learning approaches.
We find our algorithm, which significantly extends existing work by various features, to perform very well.
Our machine-learning trained decision tree surprises with a very simple, but highly precise model. 
We apply our classifier models against a large-scale measurement and find different, but always significant, counts of siblings based on domain lists. 
We discuss outliers, likely caused by proxies or middleboxes, and the influence of leap seconds onto the TCP timestamping clock. 
We release our ground truth, code and data to the scientific community to allow for reproducibility and further research in this area.

\textbf{Future Work:}
One direction of future work is the curation of larger sibling ground truth \dss. 
We hope to start this process with the release of our \gt \ds on GitHub. 
Another direction is the reduction of measurement effort in terms of duration, frequency, or both. Especially, the very discriminative raw TCP timestamp feature should work well with few data points, and hence only require few packets instead of hour-long measurements.
Furthermore, the application of our technique on passive traffic captures to distinguish siblings sounds like a promising research goal.

\textbf{Data Release: }%
We publish our curated \gt \ds, acquired raw data, and developed source code for both reproducibility (cf.~\cite{AcmArtifacts,reproduc2017}) and use by other researchers under:
\centerline{\texttt{\url{https://github.com/tumi8/siblings}}}
This website includes directions on how to obtain the large raw \ds  from an archival storage server. 

\textbf{Acknowledgments: }
We gratefully thank the various contributors of ground-truth server data.
This work has been supported by the German Federal Ministry of Education
and Research, project X-CHECK, grant 16KIS0530, and project AutoMon, grant 16KIS0411.

\apptocmd{\sloppy}{\hbadness 10000\relax}{}{}
\apptocmd{\thebibliography}{\raggedright}{}{}
\bibliography{siblings}
\bibliographystyle{abbrv}
\end{document}